# DESIGNING III-V MULTIJUNCTION SOLAR CELLS ON SILICON

J. P. Connolly [1], D. Mencaraglia [2], C. Renard [3], D. Bouchier [3]

[1] Universidad Politécnica de Valencia, NTC, B 8F, 2º. Camino de Vera s/n. 46022, Valencia, Spain
[2] LGEP, UMRCNRS 8507, Supelec, Université Pierre et Marie Curie, Université Paris-sud,
11 rue Joliot-Curie, 91192 Gif-sur-Yvette, France
[3] IEF, UMR CNRS 8622, Université Paris-Sud, Orsay F-91405, France

ABSTRACT: Single junction Si solar cells dominate photovoltaics but are close to their efficiency limits. This paper presents ideal limiting efficiencies for tandem and triple junction multijunction solar cells subject only to the constraint of the Si bandgap and therefore recommending optimum cell structures departing from the single junction ideal. The use of III-V materials is considered, using a novel growth method capable of yielding low defect density III-V layers on Si. In order to evaluate the real potential of these proposed multijunction designs, a quantitative model is presented, the strength of which is the joint modelling of external quantum efficiency and current-voltage characteristics using the same parameters. The method yields a single parameter fit in terms of the Shockley-Read-Hall lifetime. This model is validated by fitting experimental data of external quantum efficiency, dark current, and conversion efficiency of world record tandem and triple junction cells under terrestrial solar spectra without concentration. We apply this quantitative model to the design of tandem and triple junction solar cells, yielding cell designs capable of reaching efficiencies without concentration of 32% for the best tandem cell and 36% for the best triple junction cell. This demonstrates that efficiencies within a few percent of world records are realistically achievable without the use of concentrating optics, with growth methods being developed for multijunction cells combining III-V and Si materials.

1. INTRODUCTION

Multijunction cells remain the most successful high efficiency design concept, and the only one successfully commercialised. Nevertheless, these structures are still restricted to niche applications in space and under concentration, although generic terrestrial applications are tantalisingly close. The materials most suitable for multijunction designs are the III-V materials families, given their range of optical and material properties where incompatibilities, especially due to lattice constants, may be managed or even eliminated. The classic example is the growth of GaInP and GaInAs top and middle gap junctions on Ge substrates as bottom cell. This triple junction has achieved efficiencies [1] under concentration greater than 40% with both lattice matched and lattice mismatched approaches.

The main obstacles to widespread use of multijunctions in renewable energy supply are materials and fabrication issues. The materials cost is partly due to the scarcity of some essential III-V materials such as In and Ga, but more importantly, it is due to the use of expensive substrates, since even Ge remains relatively costly in the competitive photovoltaic market.

Concerning cheap substrates, the clear front-runner is Si, which brings with it the advantage of low cost industry-standard CMOS processing technologies, and more specifically brings the advantage of the most advanced development for commercial solar cells. The additional need for III-V materials is barely a few microns, reducing the materials cost to acceptable levels.

However, Si suffers from a lack of semiconductors lattice matched to it, and from an indirect bandgap and low absorption coefficient. This is an obstacle for most opto-electronic applications including multijunction solar cell design. Although all-Si structures have been proposed, using amorphous silicon and silicon-based metamaterials to fabricate multijunction cells [2], the most attractive and most flexible option remains growth of high quality III-V materials on Si.

An early attempt to address this issue by Yang [3] simply grew a thick AlGaAs buffer layer on an active Si substrate, with a simple three-terminal design. This approach did not exceed efficiencies of 20%, and has not to date been developed further. Some years later, Taguchi et al. [4] adopted a not dissimilar structure with a high quality GaAs cell grown on a GaAs substrate and transferred to a Si substrate by liftoff. They chose a four-terminal design but again failed to demonstrate more than 19% efficiency.

Geisz and co-workers have since reported results on two-terminal designs. They have investigated both lattice matched attempts using nitrides [5] and lattice mismatched approaches using graded buffer techniques for strain relaxation minimising bulk defect densities [6]. The work of Lueck et al. [7] in the same year demonstrated similar techniques and achieved an efficiency of 17% under AM1.5G spectrum. Work in this field continues with Putyato et al. [8] developing graded buffer approaches. While these methods have made progress, the efficiencies achieved remain subject to materials quality limitations due to the fundamental problem of high defect densities, despite the use of buffers to reduce them to acceptable levels.

This paper presents a quantitative study of silicon-based multijunction solar cells for terrestrial applications developed by the Multispectral Solar Cells on Silicon (MULTISOLSI) project [9]. The study assumes an AM1.5G spectrum which corresponds to photovoltaic systems with no solar concentration. This is suitable for strategies prioritising a low system cost over cell efficiency and peak power. However, we are also aiming at relatively low cell cost with our silicon-based design, while nevertheless reaching high efficiencies. While a detailed cost analysis on this point is beyond the scope of this paper, we note that the cell design methodology we describe is equally applicable to concentrating photovoltaics under direct solar spectra. The choice of a global spectrum does not, therefore, limit the scope of the design methods we will describe.

A detailed growth investigation presented in a companion paper [10] reports on progress in the development of III-V growth on Si using novel low cost three dimensional growth techniques within this project. This technique involves epitaxial lateral overgrowth (ELO) of lattice mismatched polar semiconductors on Si via growth of nano-seeds in apertures opened in thin $SiO_2$ layers. This technique has demonstrated defect free growth and an absence of antiphase domain formation [11] due to the small initial growth area. This paper therefore does not consider the formidable materials issues which remain to be resolved within MULTISOLSI, but concentrates on the design stage only by evaluating the potential of III-V structures on Si using a validated, quantitative model.

We first examine the suitability of silicon for multijunction cells from the commonly used ideal theoretical viewpoint of the radiative limit, but amended to maximise the radiative efficiency of multijunction cells with non-ideal bandgaps and layer thicknesses. We then describe a quantitative analytical model capable of accurately modelling record multijunction cells. On this basis we present the design basis for dual and triple junction cells being currently developed in the MULTISOLSI project.

## 2. IDEAL EFFICIENCY LIMITS

The objective of this study being cheap terrestrial multijunction photovoltaic systems, it is instructive to first look at the suitability of Si in the ideal limit under the AM1.5G spectrum which we do by looking at the maximum efficiencies achievable using silicon in multijunction cells in the radiative efficiency limit, using the form published by Henry [12]. The ideal system considered consists of a stack of $N$ subcells of index $i$, each of unit external quantum efficiency (EQE), and converting every photon absorbed into a current carrier. Each subcell of bandgap $E_{gi}$ absorbs light in the wavelength range corresponding to energies between $E_{gi}$, and $E_{gi+1}$ with obvious modifications for the first and $Nth$ cells. For this system, under spectral irradiance $F$, and adopting the photovoltaic sign convention of positive photocurrent, the short circuit current density $J_{SCi}$, and the radiative saturation current density $J_{0i}$ of each subcell take the form

$$J_{sc_i} = q \int_{E_{g_i}}^{E_{g_i+1}} F(\lambda) \frac{\lambda}{hc} d\lambda \quad (1)$$

$$J_{0i} = \frac{q(n_i^2+1) E_{gi}^2 K_B T}{4\pi^2 \hbar^3 c^2} e^{\frac{-E_{gi}}{K_B T}} \quad (2)$$

where $n_i$ is the subcell refractive index (taken as 3.5 here as in Henry's work [12] which is representative of relevant semiconductors, and in particular of Si), and where the other symbols have their usual meaning. For series connected subcells, again with the positive photocurrent sign convention, the overall multijunction bias $V$ as a function of multijunction series current density $J$ is given by

$$V(J) = \frac{K_B T}{q} \ln \sum_1^N \left( \frac{J_{sc_i} - J}{J_{0i}} + 1 \right) \quad (3)$$

thereby defining the $J(V)$ current-voltage characteristic. The maximum of the corresponding power – voltage characteristic yields the maximum possible efficiency achievable for any structure of $N$ ideal junctions. This is usually presented as a plot of efficiency as a function of band-gap showing iso-efficiency contours which allows the potential of materials for high efficiency to be estimated at a glance. On figure 1a, we see the maximum efficiency achievable with two gaps, which is 42.2% in the conditions specified above for materials with bandgaps 0.96eV and 1.64eV.

Replacing the lower 0.96eV material with slightly higher band-gap silicon has a very slightly lower efficiency limit of 41.9% for bandgaps 1.12 (Si) and 1.74eV. The good match is underlined by the fact that the two solutions occupy two local neighbouring maxima on the iso-efficiency contour.

This bandgap restriction may be relaxed by considering subcells with less than unit EQE, specifically with optically thinned subcells in mind. While this reduces the limiting efficiency achievable due to increased thermalisation, it may be useful in order to mitigate material constraints. Figure 1b shows the efficiency obtainable in the radiative limit by thinning the top cell such that the ideal bottom cell gap approaches Si. Having achieved this, the top cell ideal bandgap is one with 68% overall absorptivity, consequently transmitting 32% of the spectrum to the bottom Si. The optimum top cell bandgap of 1.46eV in this case being fortuitously close to GaAs at 1.42eV is an accidental bonus as we shall see subsequently.

Figure 2a shows similar evaluations of Si for triple junction cells, where the ideal limit has a middle junction ideal bandgap of 1.21eV for an efficiency of 47.2%. Figure 2b shows the efficiency profile for the maximum achievable with a silicon middle bandgap cell. We see that a slightly lower efficiency limit of 45.4% is achievable with a silicon middle gap cell. Although the loss is greater than for the tandem case, this retains record breaking potential.

These introductory remarks set out the suitability of silicon for multijunction cells in the ideal limit. In the following sections we go on to describe a realistic, quantitative model of triple junction cells in order to design multiple junction cells on silicon substrates.

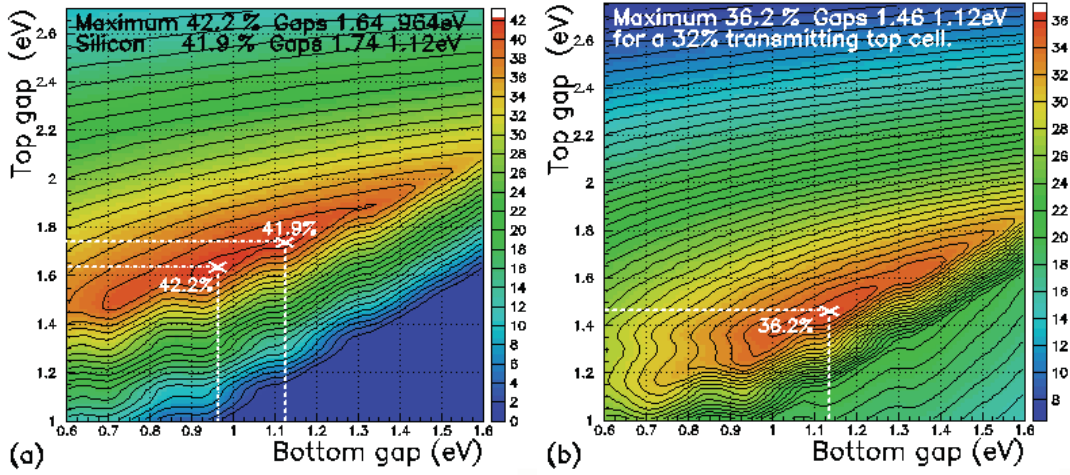

**Figure 1** Tandem junction radiative efficiencies for an AM1.5G solar spectrum showing **(a)** the maximum dual junction efficiency of 42.2% (bandgaps 1.64, 0.964eV) and the silicon based tandem efficiency maximum at 41.9% (bandgaps 1.74, 1.12eV), and **(b)** maximum efficiency achievable with thinning the top cell (32% optical transmission) for current matching with a Si 1.12eV bottom cell.

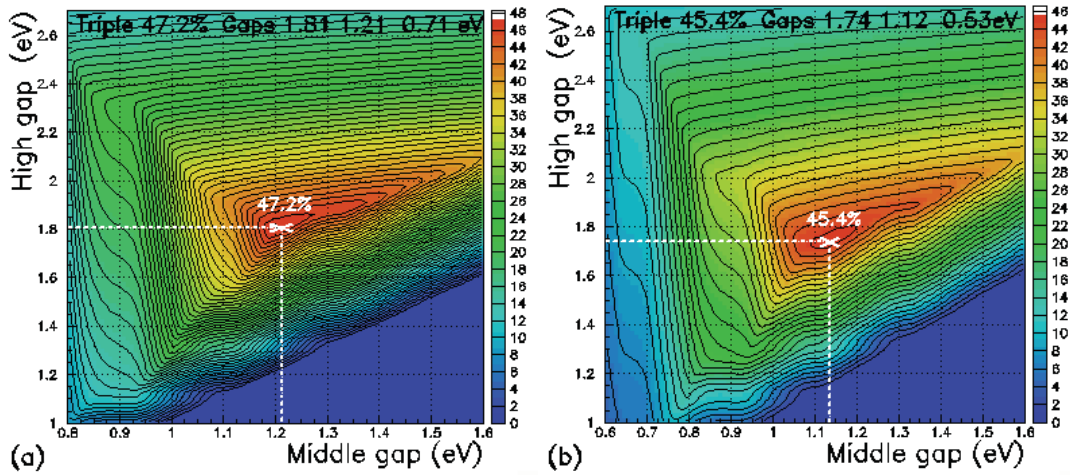

**Figure 2** Triple junction solar cell radiative limit efficiencies for an AM1.5G solar spectrum as a function of top and middle bandgap subcells for a single lower bandgap cell in each case, showing **(a)** the absolute maximum of 47.2% (bandgaps 1.81, 1.21 and 0.71eV) and **(b)** the maximum of 45.4% occurring for a silicon middle gap cell (bandgaps 1.74, 1,12, 0.53eV).

## 3. QUANTITATIVE MULTIJUNCTION MODEL

In order to design test structures for the novel growth method, we use an analytical model SOL which has been reported in more detail elsewhere [13], but of which we sketch the main aspects here firstly for a single junction. With material parameters for the majority of semiconductors in the III-V family, drawn from the literature, including Si(Ge), the model evaluates the external quantum efficiency (EQE) and photocurrent $J_{PH}$ as a function of bias by standard analytical solutions [12] of transport and continuity equations in the depletion approximation for one-dimensional structures with abrupt interfaces. The point of interest in this model is the detailed accounting of loss mechanisms by explicit solution of non-radiative and radiative recombination losses in the different regions of the cell, and the inclusion of series and shunt resistances for each subcell.

As such, the radiative and non-radiative recombination currents in the charge-neutral layers are described by the Shockley diffusion or injection current [14] as follows

$$J_S(V) = q\, e^{\frac{qV}{K_bT}} - 1 \left[ \begin{array}{c} \frac{n_{ip}^2}{N_A}\frac{D_n}{L_n} \frac{\frac{S_n L_n}{D_n}\cosh\frac{x_p}{L_n}+\sinh\frac{x_p}{L_n}}{\frac{S_n L_n}{D_n}\cosh\frac{x_p}{L_n}+\sinh\frac{x_p}{L_n}} \\ + \frac{n_{in}^2}{N_D}\frac{D_p}{L_p} \frac{\frac{S_p L_p}{D_p}\cosh\frac{x_n}{L_p}+\sinh\frac{x_n}{L_p}}{\frac{S_p L_p}{D_p}\cosh\frac{x_n}{L_p}+\sinh\frac{x_n}{L_p}} \end{array} \right]$$
(4)

where $n_{ip}$ is the intrinsic carrier concentration in the p layer doped at a level $N_A$, of surface recombination velocity $S_n$, and corresponding parameters $n_{in}$ and $N_D$ in the n doped later with its recombination velocity $S_p$.

The non-radiative recombination currents in the space-charge region are described by the Shockley-Read-Hall recombination current (SRH, [15]) with the standard expression:

$$J_{SRH}(V) = q \int_{x_1}^{x_2} \frac{p(x)n(x) - n_i^2}{\tau_n(p(x) + p_t) + \tau_p(n(x) + n_t)} dx \quad (5)$$

where $x_1$ and $x_2$ are the edges of the space-charge region, $n(x)$ and $p(x)$ are electron and hole concentration as a function of position, $n_t$ and $p_t$, the electron and hole trap occupation densities for mid-gap trap levels, and electron and hole non radiative lifetimes $\tau_n$ and $\tau_p$, respectively.

The radiative recombination currents in space-charge layers are evaluated from the absorption coefficients and quasi-Fermi level separation in each region of the cell as

$$J_{RAD}(V) = q \int_0^\infty \frac{2n^2}{h^3 c^2} \frac{E^2}{e^{(E-q\varphi)/K_b T}} \int_S \alpha(E,\theta,S) dS \, dE \quad (6)$$

where $n$ is the refractive index of the material, $\Delta\phi$ is the quasi-Fermi level separation, and the other symbols have their usual meanings. The absorptivity $\alpha(E,\theta,S)$ is the line integral over position through the different layers of the cell along the optical path of radiation at angle $\theta$ with the normal exiting or entering surface $S$, the total emitting surface in three dimensions [16].

The sum of these three recombination currents therefore describes radiative and non-radiative recombination dark currents in the cell. Following this analytical method, a little more analysis can evaluate the radiative efficiency of the structures [13].

A noteworthy feature of this approach combining the modelling of radiative and non-radiative currents together with the EQE is that the all the parameters determining these currents except for the SRH lifetimes are determined by the EQE, thereby minimising the number of free parameters. The combination of dark current and EQE fitting therefore leaves as only free parameters the SRH lifetimes. In absence of better knowledge, and consistently with current continuity, electron and hole non radiative SRH lifetimes are asumed equal in practice, resulting in a single free parameter: The SRH lifetime.

The efficiency of the cell is then evaluated from the light current $J_L$ assuming superposition of dark current and photocurrent $J_{PH}$ as

$$J_L(V) = J_{PH} - (J_S + J_{SRH} + J_{RAD}) \quad (7)$$

by finding the maximum power point on the light current curve.

The corresponding multijunction structure requires series connection of subcells via tunnel junctions. At this design stage of the MULTISOLSI project, we rely on published values as a guideline and on numerical calculations for Si tunnel junctions we have performed with SILVACO software. These show that in GaAs tunnel junctions [17], for example, peak tunelling currents above $10^4$ A.m$^{-2}$ are readily obtainable. The total current density obtainable with the global AM1.5G terrestrial spectrum is 726 A.m$^{-2}$. The multijunction current flow is therefore at least an order of magnitude less than that of a suitable GaAs tunnel junction, which therefore serves to cause a small voltage loss, as it is designed to. As such, in this work we treat the tunnel junctions as absorbing layers from an optical point of view, and as series resistance elements from the electrical point of view. In addition, we note that the losses incurred due to the tunnel junctions are of the order of 0.1% absolute due to a combination of optical and electrical penalties.

The multijunction current-voltage characteristic is therefore evaluated by separately calculating the current-voltage characteristics of the sub-cells optically and electrically connected in series, including parallel and series resistance losses and numerically evaluating the resulting multijunction solar cell current. This approach has shown itself capable of quantitatively reproducing world record EQE, dark current, and light current characteristics of tandem and triple junction cells published by Japan Energy Corp. [18] and Spectrolab [19] (tables 1 and 2). While complete details including fits of EQE, dark, and light current voltage characteristics are available elsewhere [13], here we show the good fit to the EQE of the triple junction cell (figure 3a). As described in detail in ref. [13], the modelling in both cases relies on minority carrier transport as a function of doping given in the literature.

It is important to note that the experimental EQE of this Spectrolab triple junction record cell is close to 100%. This in part due to the very low reflectivity achievable with multi-layer AR coats, such as the double-layer AR coat used by the modelling here. The high EQE is also due to the relaxation of minority carrier transport constraints in multijunction cells, which is a consequence of each subcell in the stack needing to absorb a shorter range of incident wavelengths close to the band-edge. In particular, the impact of front surface recombination is reduced.

The corresponding single-parameter dark current fitting is shown in figure 3b. It is worth underlining that this dark current fit uses the transport parameters validated by the EQE fitting as we have discussed, with the single SRH lifetime in the space-charge region the only fitting parameter for the dark current fit. The strength of this approach is illustrated by the close dark current fit, shown in figure 3b, to available data for the triple junction sub-cells, in terms of radiative and non-radiative recombination mechanisms.

This combined validation of transport by fitting EQE and dark current gives confidence in the transport parameters used, and confirms that the very high EQE predicted is to some degree over-optimistic only in the analytical calculation of the double-layer AR coat reflectivity. We cannot avoid this slight over-estimation, however, without arbitrarily reducing cell transport efficiency below published values, or by including external parasitic losses which are present in real AR coats, such as absorption losses due to to impurities or back-scattering losses due to surface roughness for example. This point is further confirmed by comparison with the EQEs achieved in practise by Spectrolab which also nearly reach 100%, as shown in figure 3a.

The principal transport parameters of the EQE and dark current fits are summarised in table 3 for the JEC tandem and Spectrolab triple cell fits. The quoted diffusion lengths in both cases are obtained from the literature as stated earlier for each emitter and base regions in each of the subcells in both cases. The diffusion lengths in all cases are of the order of microns, with the exception of Ge which benefits from significantly longer diffusion lengths as experimental studies in the literature have shown.

Also indicated in table 3 are surface recombination velocities. The impact of these on the EQE cannot be easily decoupled from the diffusion lengths without a numerical fitting parameter exploiting the different wavelength dependence of recombination velocities and diffusion lengths. Here we use a manual method and, using a fixed diffusion length determined by the material composition and doping, we vary the front and back recombination lengths to fit the short wavelength and long wavelength response respectively. Following this procedure, these transport parameters are fixed by the EQE fit and determine the Shockley injection and surface recombination dark current mechanisms.

The EQE calculation therefore fixes the transport parameters. As a result, the only fitting parameter for the dark current calculation is the Shockley-Read-Hall lifetime in the space-charge region, given in table 3. The values in all cases, including the much longer SRH liftime for the Ge subcell, are determined from the available high bias data available from the Spectrolab publication [19]. The trends observed as a function of material, and in particular the long SRH lifetime we see for Ge, are confirmed by a brief review of net SRH lifetimes in the literature. Somewhat fortuitously, the data available, while not supplying separate IV data for the Ge subcell, provides data for the middle and top band-gap subcells, in a bias range that allows us to fit the total triple junction dark current, and therefore the Ge subcell dark current, with a high degree of confidence which results from the exact simultaneous fit obtained for middle GaInAs subcell, top GaInP subcell, and the combined triple dark currents seen in figure 3b.

Summarising this modelling, tables 1 and 2 show the main cell performance figures of merit under AM1.5G in the cases of both the Japan Energy Corp. tandem record [18] and the Spectrolab triple junction record [19], showing good agreement with in the worst case a ≈ 3% relative under-estimation of efficiency by SOL in the case of the tandem cell.

In the following sections we will use this modelling approach to look at multijunction designs combining III-V materials on silicon. We will assume that material quality equal to that shown in these record cells is achievable using the low defect density III-V on silicon growth methods we have mentioned [9,10].

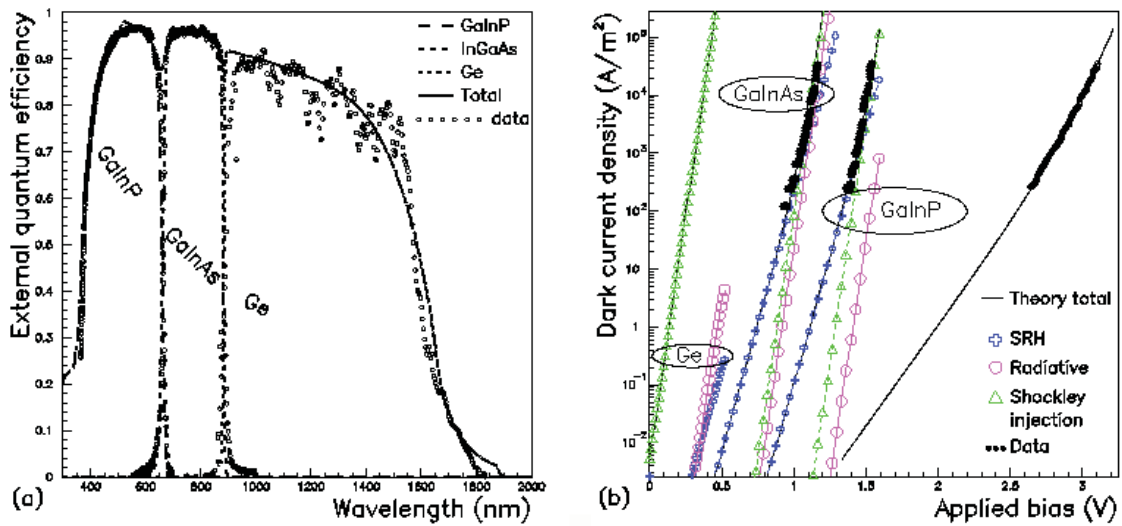

**Figure 3** Record triple junction [19] showing (a) EQE data and modelling with calculated reflectivity and (b) dark current modelling in terms of radiative and non-radiative contributions together with available data.

|  | Jsc (A/m2) | Voc (V) | FF (%) | Efficiency (%) |
|---|---|---|---|---|
| **Japan Energy Corp.** | 142.5 | 2.49 | 85.6 | 30.3 |
| **Model SOL** | 139.5 | 2.32 | 87.0 | 29.4 |

**Table 1**. Experimental data and SOL modelling of a record tandem solar cell [18] under an AM1.5G spectrum.

|  | Jsc (A/m2) | Voc (V) | FF(%) | Efficiency (%) |
|---|---|---|---|---|
| **Spectrolab** | 143.7 | 2.62 | 85 | 32.0 |
| **Model SOL** | 143.2 | 2.62 | 86.0 | 32.4 |

**Table 2.** Experimental data and SOL modelling of a Spectrolab triple junction record cell [19] under an AM1.5G spectrum.

| *Tandem and triple cell transport parameters* | Ln (µm) | Sn (m/s) | Lp (µm) | Sp (m/s) | $\tau_{SRH}$ (ns) |
|---|---|---|---|---|---|
| *JEC-tandem - **GaInP*** | 0.4 | 750 | 2.5 | 40 | |
| *JEC-tandem - **GaAs*** | 1.1 | 2420 | 7.6 | 20 | |
| *Spectrolab-triple - **GaInP*** | 5.1 | 1100 | 1.5 | 35 | 50 |
| *Spectrolab-triple - **GaInAs*** | 0.28 | 3000 | 1.2 | 40 | |
| *Spectrolab-triple - **Ge*** | 62.3 | 140 | 499 | 5 | $10^3$ |

**Table 3.** Transport parameters: p-doped layer electron diffusion length diffusion lengths **Ln** and n-doped layer hole diffusion length **Lp**, electron and hole recombination velocities Sn and Sp determine the EQE as discussed in the text. The dark current is determined by the same transport parameters plus $\tau_{SRH}$ the Shockley-Read-Hall (SRH) lifetime which is the single free parameter.

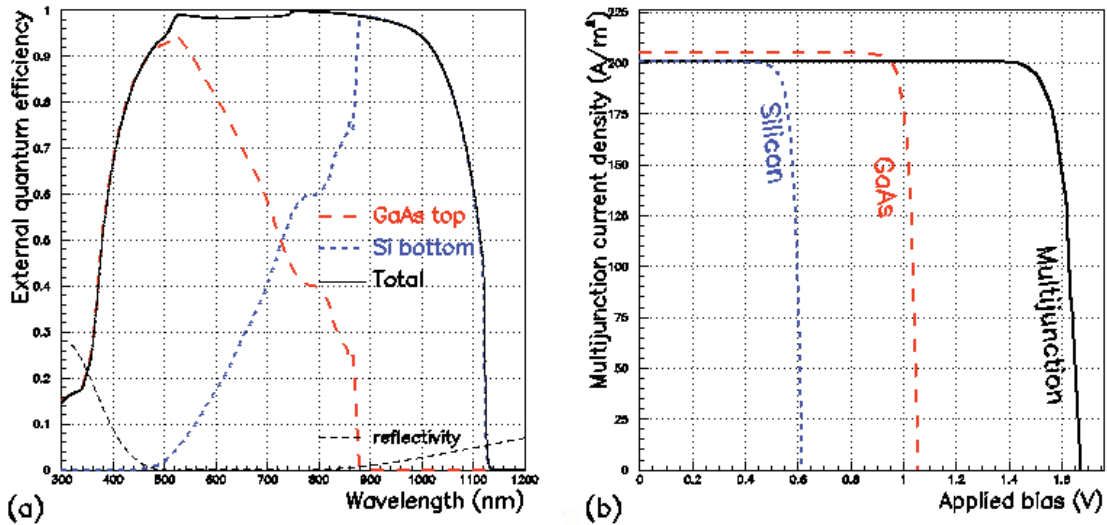

**Figure 4** Modelled EQE (a) of the silicon based tandem structure with a cross optimised AR coat and GaAs high gap subcell thickness to ensure current continuity. The modelled light current characteristics (b) show the close current matching achieved, and the lack of significant series resistance associated with the tunnel junction.

| Layer | Material | Thickness (μm) | Doping (m$^{-3}$) |
|---|---|---|---|
| *MgF2 (110nm)/ ZnS (65.2nm) antireflection coat optimised at 800nm* | | | |
| Window | AlGaAs | 0.02 | p-type C > 10$^{25}$ |
| Base | GaAs | 0.15 | p-type C=2x10$^{24}$ |
| Emitter | GaAs | 0.2 | n-type Si=2x10$^{23}$ |
| Tunnel | GaAs | 0.01 | n-type Si degenerate |
| Tunnel | GaAs | 0.01 | p-type C degenerate |
| Emitter | Si | 5 | p-type B=10$^{24}$ |
| Base | Si | 140 | n-type P=10$^{21}$ |
| BSF | Si | 1 | n-type P=10$^{23}$ |

**Table 4** GaAs on Si substrate tandem structure with a suitable joint optimisation of GaAs layer thickness and dual layer AR coat design.

| $J_{SC}$ (A.m$^{-2}$) | $V_{OC}$ (V) | $V_{MP}$ (V) | FF (%) | Efficiency (%) |
|---|---|---|---|---|
| 200.4 | 1.67 | 1.50 | 87 | 29.2 |

**Table 5** GaAs/Si tandem cell performance under an AM1.5G spectrum.

## 4. MULTIJUNCTION STRUCTURES ON SILICON

We have identified ideal bandgaps for tandem and triple junction cells based on silicon and described a quantitative model. We use this to now describe the prototypes for fabrication by the novel growth method being developed within MULTISOLSI.

Since at present only binary III-V crystalline growth without propagation of bulk defects has been demonstrated, we first restrict the discussion to binary compounds for tandem structures, before discussing the extension to ternary compounds for tandem and triple junction solar cells.

### 4.1 Binary tandem

The first structure we propose is a GaAs on Si structure with a dual layer MgF$_2$/ZnS anti-reflection (AR) coat. We again note that the AR coat is idealised in the sense that it represents in all cases the best possible case with no parasitic losses, such as through absorption in the AR coat layers.

The tandem limiting efficiency calculations (figure 1) show that the optimal top cell bandgap is 1.74eV for a Si lower gap cell, which is much greater than the 1.42eV bandgap of GaAs. This translates as a significant current mismatch with the GaAs cell current limited by the Si low gap cell.

As mentioned in our discussion of ideal radiative limits, this may be much improved however by adjusting the current balance between the two cells by adjusting the AR coat on the real cell and by adjusting the layer thicknesses in consequence. We found that the limiting efficiency of just 20.4% for an opaque GaAs cell on Si could be enormously increased to slightly more than 36% by thinning the top cell such that it absorbs only 68% of the incident photons with energies above its bandgap. While significantly lower than the limiting efficiency of 42.2% (figure 1a), this is nonetheless an impressive target for a tandem solar cell, and a suitable test-bed for the growth methods being developed within the MULTISOLSI project.

For our binary semiconductor tandem design, we therefore adopt the strategy of thinning a GaAs cell in order to address this issue. The structure optimisation consists of jointly optimising the AR coat optimum wavelength and the GaAs cell thickness.

The substrate optimisation is independent and simpler, consisting solely of the standard optimisation of its EQE in terms of the trade-off between minority carrier collection efficiency versus total absorptivity for this layer. This optimisation proposes a Si substrate for this structure of 140μm total thickness with a shallow highly doped p-layer in contact with the GaAs tunnel junction.

The resulting structure is shown in table 4. This first proposed test for the novel low defect density growth method [1] yields the EQE shown in figure 4. The efficiency achieved in AM1.5G and therefore without concentration is 29% (table 5), which compares favourably with the long-standing 30% record result achieved by Takamoto and co-workers of the Japan Energy Corp. [18] in 1997 for the GaInP/GaAs tandem cell under an AM1.5G terrestrial spectrum which we discussed in the modelling section.

4.2 Ternary tandem

Lattice matched $Ga_{0.51}InP$ is a natural choice for a multijunction cell as it can be grown on a binary GaAs layer and because compatible windows, of essentially lattice matched AlGaAs or preferably AlInP exist. However its bandgap of 1.87eV, being significantly higher than the ideal 1.74eV, leads to this subcell being severly current limiting. The highest efficiency achievable is therefore just 30.16% which is barely better than the GaAs tandem.

Considering the GaInP top cells without the advantages of lattice matching to GaAs, the use of lower P compositions do enable current matching. While this in principle leads to efficiencies reaching over 32% for $Ga_{0.34}InP$, the absence of suitable lattice matched window materials makes this an impractical choice.

The only suitable ternary remaining is GaAsP. This benefits from a lattice matched AlInP window layer for a range of compositions [20]. Figure 5a shows the EQE of a GaAsP based triple junction cell with an optimum AR coat of minimum reflectivity 700nm, which yields a good current match as illustrated by the light IV shown in figure 5b.

This structure is closer to the ideal limit in that no compromise of optically thinning the top cell is needed. The optimisation of bandgap and AR coat delivers a composition of $GaAsP_{0.22}$ with a bandgap of 1.68eV. As expected, this is close to radiative limit value of 1.74eV. The not insignificant difference is due to the visibly non-ideal response of $GaAsP_{0.22}$ near the band-edge, which is due mainly to the non-ideal minority carrier transport in this material. This results in optimal emitter and base layer thicknesses being less than perfectly opaque, and therefore in the $GaAsP_{0.22}$ top cell slightly under-producing photocurrent. As a result, the quantitative optimisation procedure yields a slightly lower bandgap than the ideal.

The reduction in corresponding efficiency is due to the non-ideality of the material primarily from the remaining partial transparency near the band-edge, and from the departure of the recombination mechanisms from the radiative limit. We note in passing that evaluation of the dark current contributions (not shown for brevity) show that the overall structure is dominated by Shockley injection currents at the maximum power point.

The consequence of the improved design flexibility is that improved current matching shown in table 6. We see that this structure exceeds 32% efficiency, with higher Jsc, $V_{OC}$ and 3% absolute increase in efficiency with respect to the thinned GaAs tandem.

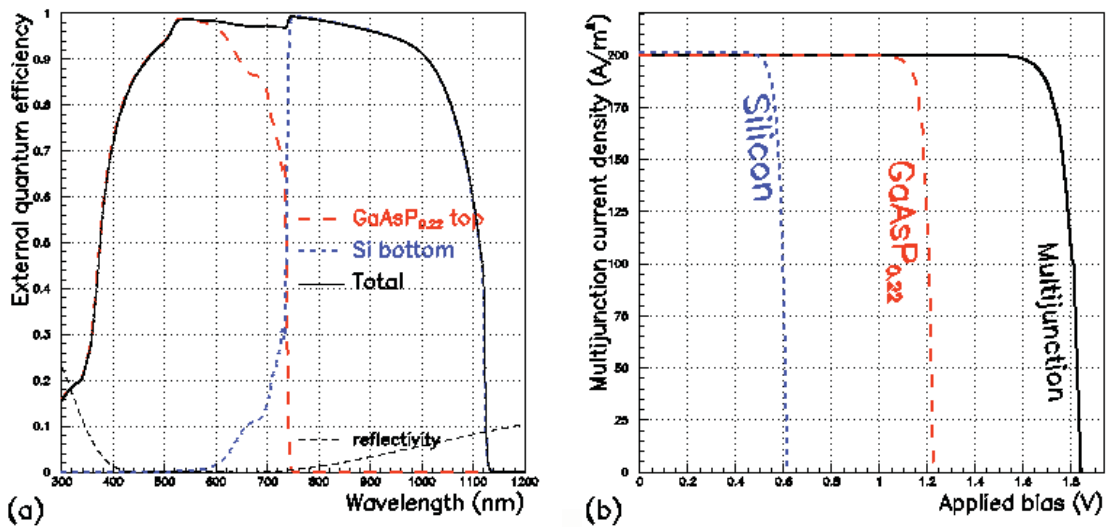

**Figure 5** Ternary $GaAsP_{0.22}$ / Si tandem response modelling. The EQE (a) shows a less than ideal top cell EQE near the band-edge resulting in an optimum top-cell gap (1.68eV) slightly lower than the ideal limit (1.74eV) which nevertheless results (b) in close current matching and an efficiency of over 32%.

| $J_{SC}$ (A m$^{-2}$) | $V_{OC}$ (V) | $V_{MP}$ (V) | FF (%) | Efficiency (%) |
|---|---|---|---|---|
| 199.7 | 1.84 | 1.65 | 88 | 32.2 |

**Table 6** GaAsP/Si tandem cell performance under an AM1.5G spectrum.

### 4.3 Triple junction structures

The extension of the tandem to triple junction devices is obvious, but requires epitaxy on both sides of the Si substrate which thereby takes on the role of middle junction solar cell. The low bandgap materials available include Ge. This material is however ill suited of the growth methods we are considering because of the necessity of growing a Ge sub-cell of hundreds of microns in thickness, due to its low absorption coefficient. This leaves us with GaInAs as the only available low band-gap material. We will not consider the nitride alloys here due to challenging materials issues which remain unresolved [5].

In the following sections we will therefore rely on a GaInAs bottom cell. A longer treatment would entail fully optimising top and bottom subcell structures. In this case, for illustrative purposes, we only consider a GaInAs structure with a bandgap corresponding to the ideal limit for a Si based triple junction (fig. 2b). The bottom gap subcell structure in all cases is therefore a Ga$_{0.74}$InAs *p-i-n* structure of total thickness 6μm, having the ideal bandgap of 0.53eV for the Si based triple junction design. The thickness of this cell is a little excessive but is chosen as such in order to make the GaInAs cell essentially opaque, in line with the design philosophy of this work. In a practical device at the production stage, however, light trapping techniques ranging from back-surface reflectors to front surface texturing would be required to allow us to achieve similar performance with GaInAs layers of lower thickness. We therefore present modelling of this bottom subcell with binary and ternary top cells as before.

### 4.4 Thinned GaAs triple junction

Figure 6a shows the EQE of a triple junction cell with a thinned GaAs binary top cell for current matching. The structure is obtained by varying only the GaAs cell thickness and the AR coat optimal wavelength, while leaving the Si and GaInAs subcells optimal. In keeping with the optimisation of the GaInAs structure for the triple-cell ideal limit (fig. 2b) we find the optimum GaAs thickness is unchanged from the tandem case. The AR coat optimisation however shifts to a slightly longer wavelength of 860nm.

We find (fig. 6b) that the Si cell over-produces current slightly and is forced into forwards bias, beyond it's maximum power point. The optimisation process refered to previously shows that this configuration is preferable to that in which the GaAs and Si subcells are current matched by adjusting the AR coat and the GaAs thickness, and are both forced past their maximum power points by the current limiting GaInAs bottom gap cell.

The AM1.5G efficiency achieved in this case is 32.9%, as shown in table 7.

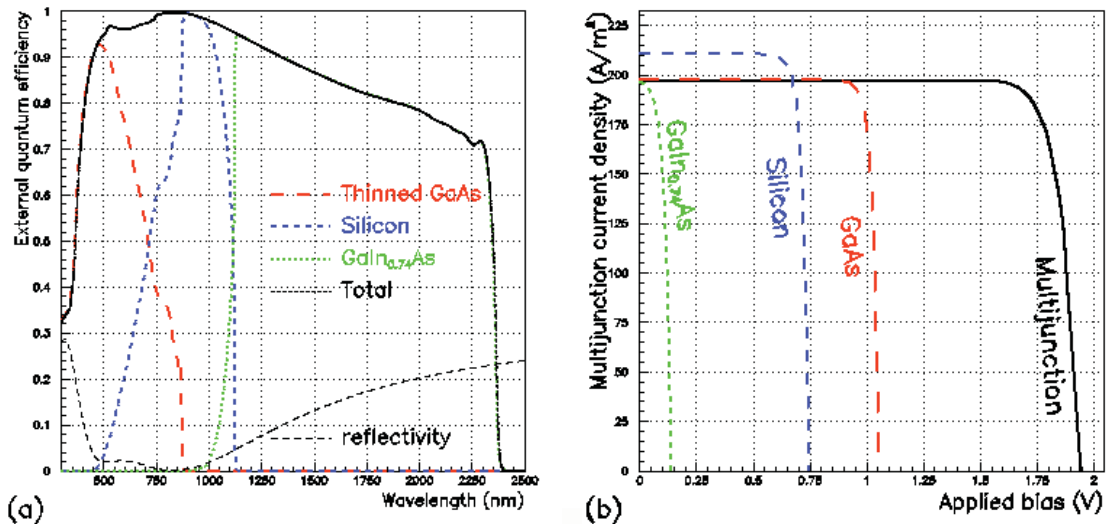

**Figure 6** Thinned GaAs - Si - GaInAs triple junction structure modelled performance with joint optimisation of GaInAs composition, GaAs thickness, and AR coat, reaching 32.9%.

| $J_{SC}$ (A m$^{-2}$) | $V_{OC}$ (V) | $V_{MP}$ (V) | FF (%) | Efficiency (%) |
|---|---|---|---|---|
| 198.8 | 1.95 | 1.69 | 85 | 32.9 |

**Table 7** Triple junction thinned GaAs / Si / GaInAs cell modelled performance under an AM1.5G spectrum.

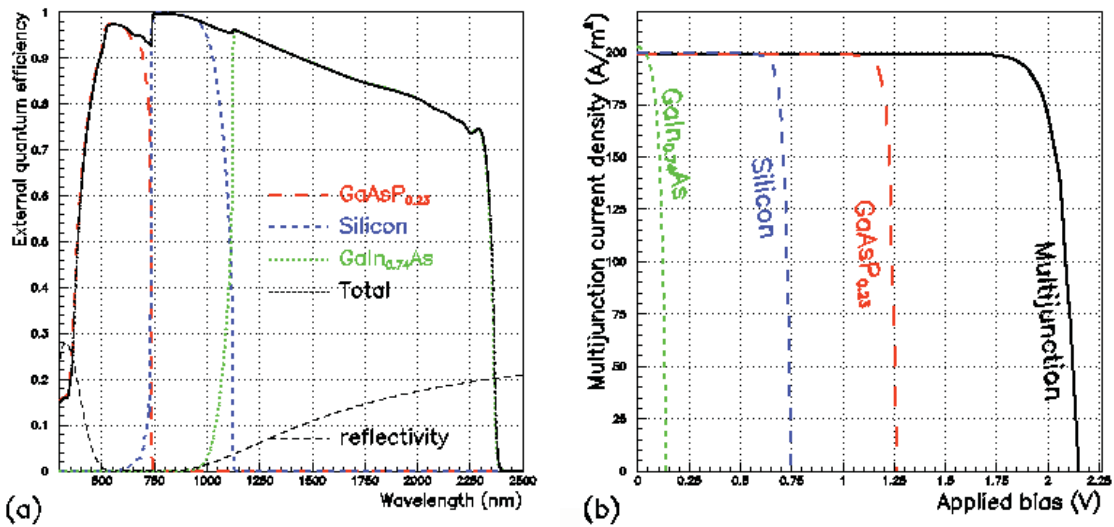

**Figure 7** Modelled performance of the triple junction cell with GaAsP top cell reaching 36.5%

| $J_{SC}$ (A m$^{-2}$) | $V_{OC}$ (V) | $V_{MP}$ (V) | FF (%) | Efficiency (%) |
|---|---|---|---|---|
| 199.0 | 2.15 | 1.90 | 85 | 36.5 |

**Table 8** Modelled performance of the triple junction GaAsP$_{0.23}$ / Si / GaIn$_{0.74}$As cell under an AM1.5G spectrum.

4.5 Ternary triple junction

For the ternary case, we consider only GaAsP, as having suitable window materials as discussed ealier. In this case, as before, no compromise is necessary given the variable gap of the top sub-cell. An optimisation of AR coat and of top cell bandgap yields a GaAsP$_{0.23}$ top-cell with a bandgap of 1.69eV, which is very slightly higher than the tandem case, and which remains significantly lower than the ideal 1.74eV.

As in the tandem case, the slightly lower band-gap than expected is consistent with the less than ideal EQE of the GaAsP cell (fig. 7a), and in particular because of the significant short-wavelength losses entailed recombination at the interface with the AlInP window. The light-current modelling (fig. 7b) however shows exact current matching achieved by the optimisation process in this case.

As a result of this more flexible and better optimised structure, the efficiency achieved (table 8) is significantly higher, reaching 36.5% in AM1.5G. This represents eighty percent of the radiative efficiency limit for a triple junction cell.

5. CONCLUSIONS

Investigating the use of Si for multijunction cells, we have looked at non-ideal designs in the ideal radiative limit which include a Si subcell. These designs show that high efficiencies are achievable by balancing competing compromises of material restrictions and non-ideal optical properties.

In order to design real cells, we have developed a quantitative analytical model. We have validated this model by quantitatively reproducing state of the art light and dark current characteristics of record tandem and triple junction solar cells. The strength of the model lies in the minimisation of free parameters via consistent modelling of EQE and dark current characteristics.

This leads to dark current fitting in terms of a single free parameter, which is the Shockley-Read-Hall non radiative lifetime in the space-charge region. With the exception of the reflectivity calculations which assume loss-free dual layer MgF / ZnS AR coats, the approach to modelling has been to use published experimental values for transport parameters, in order to maximise agreement with experimental data. The quantitative modelling of the record tandem and triple junction solar cells, together with the compatibility of transport parameters used with values in the literature give a high degree of confidence in this analytical modelling methodology. The minority carrier properties and SRH lifetimes used in the modelling are therefore reliable benchmarks for the material to be grown on Si by the growth methods developed within the MULTISOLSI project.

Applying this quantitative model to realistic solar cell designs combining Si and III-V materials, we first find that an appropriate tandem design of thinned GaAs on Si can deliver an efficiency of 29%, close to the 30% world record achieved with a GaAs substrate under a terrestrial spectrum with no concentration.

Furthermore, we find that more challenging growth of ternary materials deliver tandem efficiencies greater than 32%, and triple junction efficiencies greater than 36%.

These high efficiencies are remarkable in that they are achieved without solar concentration. They nevertheless reach efficiencies just a few percent below the maximum efficiencies achieved to date, at concentrations of 500 suns and above. This lack of concentrating optics invites the clear advantage of simple photovoltaic systems. More importantly, from the cell design point of view, it significantly relaxes the design criteria for tunnel junctions between subcells, due to the significantly reduced and less demanding current flow achievable with a terrestrial global spectrum.

This work has been carried out within the MULTISOLSI project, which has demonstrated the growth of GaAs on Si without the formation of anti-phase

<em>28th European Photovoltaic Solar Energy Conference and Exhibition</em>

domains, and without the generation of bulk defects. Work continues to develop a finer theoretical understanding of the three dimensional nature of the structures involved, of the corresponding tunnel junctions, and to optimize the growth and processing techniques required to fabricate these potentially ground-breaking designs.

ACKNOWLEDGMENTS
The Multispectral Solar cells on Silicon (MULTISOLSI) project is funded by the French Agence Nationale pour la Recherche under the program ANR PROGELEC 2011 ref. ANR-11-PRGE-0009.